\documentclass[%
 aip,
cp,  
 amsmath,amssymb,
 reprint,%
]{revtex4-2}

\usepackage{graphicx}
\usepackage{dcolumn}
\usepackage{bm}

\usepackage[utf8]{inputenc}
\usepackage[T1]{fontenc}
\usepackage{mathptmx} 

\begin{document}

\title{Probing the Anisotropic Universe with Gravitational Waves}

\author{Eoin \'O Colg\'ain} 
 \email[]{eoin@sogang.ac.kr}
\affiliation{
Center for Quantum Spacetime, Sogang University, Seoul 121-742, Korea \\
Department of Physics, Sogang University, Seoul 121-742, Korea
}

\date{\today} 

\begin{abstract}
Modern cosmology rests on the working assumption that the Universe is isotropic and homogeneous at large scales. Here, we document a number of anomalous observations pointing to an anisotropic Universe in a direction consistent with the CMB dipole. If physical, this should be confirmed by any cosmological probe with sufficient data quality. To that end, we perform the first hemispherical decomposition of bright and dark standard siren $H_0$ determinations, finding that the maximum of the $H_0$ posterior is larger in the CMB dipole direction for dark sirens, but including the bright siren posterior reverses this conclusion. As may be expected, the limited number of GW observations is consistent with an isotropic Universe, but this can change going forward. It is imperative to repeat these tests as GW sources become more numerous and better localised.  
\end{abstract}

\maketitle

\section{\label{sec:level1}Introduction}
The Cosmological Principle or Friedmann-Lema\^itre-Robertson-Walker (FLRW) paradigm is a working assumption in modern cosmology, which states that the Universe is isotropic and homogeneous at suitably large scales. Note, the local Universe is \textit{not} FLRW and it is well documented that peculiar velocities map out a bulk flow in the direction of the Shapley supercluster at a distance of $\sim 200$ Mpc \cite{Hoffman:2017ako}. As an aside, such bulk flows appear anomalously large from the perspective of the flat $\Lambda$CDM model \cite{Watkins:2008hf, Kashlinsky:2008ut, Lavaux:2008th, magoulas, Howlett:2022len}. Moreover, the bulk flows may terminate at Shapley, the Vela supercluster in the Zone of Avoidance, or they may continue beyond. At the other end of the Universe, the Cosmic Microwave Background (CMB) is for all extensive purposes consistent with FLRW, although well documented anomalies exist \cite{Schwarz:2015cma}. In particular, even after the CMB has been fixed as the rest frame of the Universe by removing the dipole ($\ell =1$ multipole), there exists a residual hemispherical power asymmetry \cite{Eriksen:2003db, Hansen:2008ym, Planck:2019evm} in the direction $(l, b) = (221^{\circ}, 20^{\circ})$ (galactic coordinates) with statistical significance $\sim 2 \sigma$. This anomaly has survived the transition from WMAP to Planck, so it is not an obvious systematic. It is worth noting that this direction is close to and shares the same hemisphere as the CMB dipole direction, $(l, b) = (264^{\circ}, 48^{\circ})$, which defines our direction of motion with respect to CMB. 

The hemispherical power asymmetry leaves us with a number of options. It is either i) a statistical fluctuation, ii) a local breakdown in isotropy that does not endanger the FLRW assumption  \cite{Gordon:2006ag, Eriksen:2007pc, Dvorkin:2007jp, Galloni:2022rgg} or iii) part of a global anisotropy. In principle, these possibilities can be disentangled by studying late Universe cosmological probes beyond our local Universe, $z \lesssim 0.05$, \textit{where it is already a fact that FLRW is broken}. To begin, nearby Type Ia supernovae (SN) confirm the local anisotropic Universe \cite{Colin:2010ds}, which is expected, and simply reinforces that SN are valid distance indicators. Beyond $z \sim 0.07$, poor sky coverage, even in the Pantheon sample \cite{Pan-STARRS1:2017jku} (see Ref. \cite{Krishnan:2021jmh}), limits the effectiveness of SN to probe isotropy, but nevertheless a residual hemispherical dipole aligned with the CMB dipole and power asymmetry exists at $\sim 1.5 \sigma$ \cite{Krishnan:2021jmh}. The observation is evident in the Hubble constant $H_0$, within the flat $\Lambda$CDM cosmology, so any ansatz is minimal, and suggestively, the feature is present at low redshifts $z \lesssim 0.07$, but visibly \textit{enhanced} by higher redshift SN \cite{Krishnan:2021jmh} (see also Refs. \cite{Cooke:2009ws, Colin:2019opb} for related observations). Suggestively, strong lensing time delay observations \cite{Wong:2019kwg, Millon:2019slk} show a similar $H_0$ trend \cite{Krishnan:2021dyb}, but even when combined with SN, the significance remains below $2 \sigma $ \cite{Krishnan:2021jmh}. 

Nevertheless, the observations do not terminate with SN. Just beyond the reach of anomalous bulk flows, scaling relations in galaxy clusters have been reported to track the CMB dipole direction with significance $\sim 5 \sigma$ \cite{Migkas:2020fza, Migkas:2021zdo} (however see \cite{Giles:2022wrd}). The observation is robust across a number of scaling relations, so X-ray absorption effects are precluded. Thus, if physical, it points to either i) an anisotropic Hubble expansion or ii) a $\sim 900$ km/s bulk flow out to $\sim 500$ Mpc or $z \sim 0.1$ \cite{Migkas:2021zdo}. Interestingly, this is already beyond the $\sim 300 \, h^{-1}$ Mpc of the original ``dark flow" claim  \cite{Kashlinsky:2008ut}. Going further beyond, potentially to $z \sim 1$, there are studies based on tests outlined in Ref. \cite{1984MNRAS.206..377E} consistently pointing to excesses in the cosmic dipole \cite{Blake:2002gx, Singal:2011dy, Rubart:2013tx, Tiwari:2015tba, Bengaly:2017slg, Singal:2021crs, Singal:2021kuu} (reviewed and reproduced in Ref. \cite{Siewert:2020krp}), which have most recently reached the $\sim 5 \sigma$ significance level with quasars (QSO) \cite{Secrest:2020has}. Tellingly, radio galaxy and QSO samples are prone to different systematics and intriguingly the direction of the excess in the dipole, $(l, b) = (238^{\circ}, 29^{\circ})$ \cite{Secrest:2020has} is close to the hemispherical power asymmetry. However, the methodology of these tests  \cite{1984MNRAS.206..377E} can  be questioned \cite{Dalang:2021ruy, Murray:2021frz} and the consensus of an excess may not be absolute \cite{Horstmann:2021jjg}. Nevertheless, since studies of the cosmic dipole are conducted in heliocentric frame, one is expected to recover the CMB dipole direction, and this is indeed the case. Once again, it is worth stressing that this is more or less the direction of the hemispherical power asymmetry \cite{Eriksen:2003db, Hansen:2008ym, Planck:2019evm}. Finally, if the excess in QSOs \cite{Secrest:2020has} is real, there is a consistent signal with significance $\sim 2 \sigma$ in QSOs \cite{Luongo:2021nqh} standardised through an empirical relation in X-ray and UV fluxes \cite{Risaliti:2015zla, Risaliti:2018reu}. 

This brings us to the crux of this note. In recent years a host of observations in the late Universe now make it less clear-cut that the Universe is isotropic. Indeed, we have a good lead on a potential anisotropy, namely a strong candidate direction, which can be tested by any emerging cosmological probe \cite{Moresco:2022phi}, provided data quality is sufficient. Arguably, one of the most exciting prospects are gravitational waves (GWs) \cite{LIGOScientific:2016aoc}, since one can bypass the distance ladder on the assumption that General Relativity is correct and there is no shortage of promising forecasts and timelines for GW determinations of $H_0$ \cite{Chen:2017rfc, Zhang:2019loq, Borhanian:2020vyr, Jin:2020hmc, Wang:2021srv, Cai:2021ooo, Jin:2021pcv, Leandro:2021qlc, Zhu:2021bpp, Liu:2021yoy, Jin:2022qnj}. These forecasts of course assume an FLRW Universe. However, if the seemingly coherent anisotropies outlined above are real, there is no guarantee of a unique $H_0$, since $H_0$ is only well defined in the FLRW context. Thus, our goal here is to perform the first analysis of bright (events with an electromagnetic (EM) counterpart) and dark (events without an EM counterpart) standard sirens \cite{Schutz:1986gp}, where the $H_0$ inferences are decomposed in hemispheres along the CMB dipole direction. The motivation for doing so is that the latter currently serves as a strong candidate for an anisotropy direction. Note, given the limited number of GW events suitable for cosmology, our goal is not to substantiate an anisotropic Universe, but in the spirit of the pioneering work Ref. \cite{LIGOScientific:2017adf}, simply to get the ball moving in that (research) direction. 

\section{Analysis}
Our intent here is to piggyback on the analysis presented in Ref. \cite{Palmese:2021mjm} (see also \cite{Finke:2021aom, LIGOScientific:2021aug, Mukherjee:2022afz} for related results) by simply decomposing the $H_0$ posteriors for different GW events in hemispheres \footnote{We thank the authors for making the posteriors available.}. We refer readers to the original publication for the details of the methodology. Ref. \cite{Palmese:2021mjm} primarily focuses on binary black hole (BH) merger events, where one must assume a cosmological model, i. e. flat $\Lambda$CDM, but ignorance of redshift of the source is handled by marginalising over the ensemble of potential host galaxies. This statistical approach is expected to lead to good constraints in the not so distant future. As is clear from Table \ref{tab:GWevents}, a number of the dark siren events are at sufficiently low redshifts that they should only be mildly cosmology dependent. However, it is worth bearing in mind that Ref. \cite{Palmese:2021mjm} assumes the flat $\Lambda$CDM model with $\Omega_m = 0.3$ and $H_0$ values in the range $20-140$ km/s/Mpc. 

But before proceeding to the results, let us digress to comment on the bright standard siren programme. It is indeed true that one can read off the luminosity distance $d_{L}(z)$ directly from the GW waveform, but turning this into a determination of the Hubble constant requires a redshift from an EM counterpart \cite{Schutz:1986gp}. Therein lies the rub, since GW170817 is currently  the only event with a confident host galaxy \cite{LIGOScientific:2017adf}. Moreover, since one requires at least one neutron star (NS) in the binary for electromagnetic emission, this restricts one to lower redshifts. This is a double-edged sword. On one hand, working at lower redshifts, $z < 0.1$, one can make statements that are less sensitive to the cosmological model, e. g. \cite{Dhawan:2020xmp}. On the other, if the event is within $\sim 200$ Mpc ($z \sim 0.05$), which is the case for GW170817, where the luminosity distance is $d_{L} \sim 40$ Mpc, one is working within a non-FLRW Universe. As are result, $H_0$ determinations can and do vary on the sky \cite{McClure:2007vv} (based on Ref.\cite{HST:2000azd}) and this explains the difficulty in determining $H_0$, assuming of course that it can be determined \footnote{One can of course measure the rate of expansion, but this does not have to be isotropic, in which case $H_0$ is ill-defined.}. In short, one expects to need a large number of bright siren events to determine $H_0$ independent of the distance ladder. Given that there is only one bright siren event with a confident EM counterpart \cite{LIGOScientific:2017adf}, further discussion at this stage is premature. 

\begin{table}[htb]
\centering
\begin{tabular}{cccc}
Event & $d_{L}(z)$ (Mpc) & Hemisphere  & Reference \\
\hline
GW170817 & $43.8^{+2.9}_{-6.9}$ & CMB dipole & \cite{LIGOScientific:2017adf} \\
GW170608 & $320^{+120}_{-110}$ & delocalised & \cite{LIGOScientific:2018mvr} \\
GW170814 & $540^{+130}_{-210}$ & opposite CMB dipole & \cite{LIGOScientific:2017ycc} \\
GW170818 & $1060^{+420}_{-380}$ & opposite CMB dipole & \cite{LIGOScientific:2018mvr} \\
GW190412 & $740^{+120}_{-130}$ & CMB dipole & \cite{LIGOScientific:2020stg} \\
GW190814 & $241^{+26}_{-26}$ & opposite CMB dipole & \cite{LIGOScientific:2020zkf} \\
S191204r & $678^{+149}_{-149}$ & delocalised & \cite{GRB1} \\
S200129m & $755^{+194}_{-194}$ & opposite CMB dipole & \cite{GRB2} \\
S200311bg & $1115^{+175}_{-175}$ & opposite CMB dipole & \cite{GRB3} 
\end{tabular}
\caption{Bright and dark standard siren events from Ref. \cite{Palmese:2021mjm}. In contrast to EM events, GW events can be poorly localised and we remove two ``delocalised" events from our analysis. The lone bright standard candle event GW170817 is noticeably closer.}
\label{tab:GWevents}
\end{table}

In practice, this makes dark siren programme more attractive, since we are recording a least a factor of ten more binary BH events than binary NS events \cite{LIGOScientific:2020ibl}. That being said, without an EM counterpart, one has to marginalise over the redshift of the potential source and this process rests on ensuring that galaxy catalogues are complete. Furthermore, if the Universe is only approximately FLRW, then in contrast to EM events, one will need better localised GW events to see beyond FLRW. As explained in Ref. \cite{Trott:2021fnx}, any absolute determination of $H_0$ may be biased. That being said, here we are not interested in determining the absolute value of $H_0$, but only the relative difference across hemispheres. 

We now turn our attention to the GW events. As is clear from Figure 1 of Ref. \cite{Palmese:2021mjm}, if one is only interested in determining the absolute value of $H_0$, then one can work with delocalised events, e. g. GW170608 \cite{LIGOScientific:2018mvr} and S191204r \cite{GRB1}, provided the events still allow good overlap with galaxy catalogues. However, delocalised events risk straddling our proposed hemisphere decomposition centered on the CMB dipole. In particular, observe that S191204r is in the directions of the Pictor, Caelum or Eridanus constellations, which are essentially at the border of our hemispheres. Therefore, to be conservative we simply drop these events and mark them as ``delocalised" in Table \ref{tab:GWevents}. 

\begin{figure}[htb]
\centering 
\begin{tabular}{cc}
\includegraphics[width=80mm]{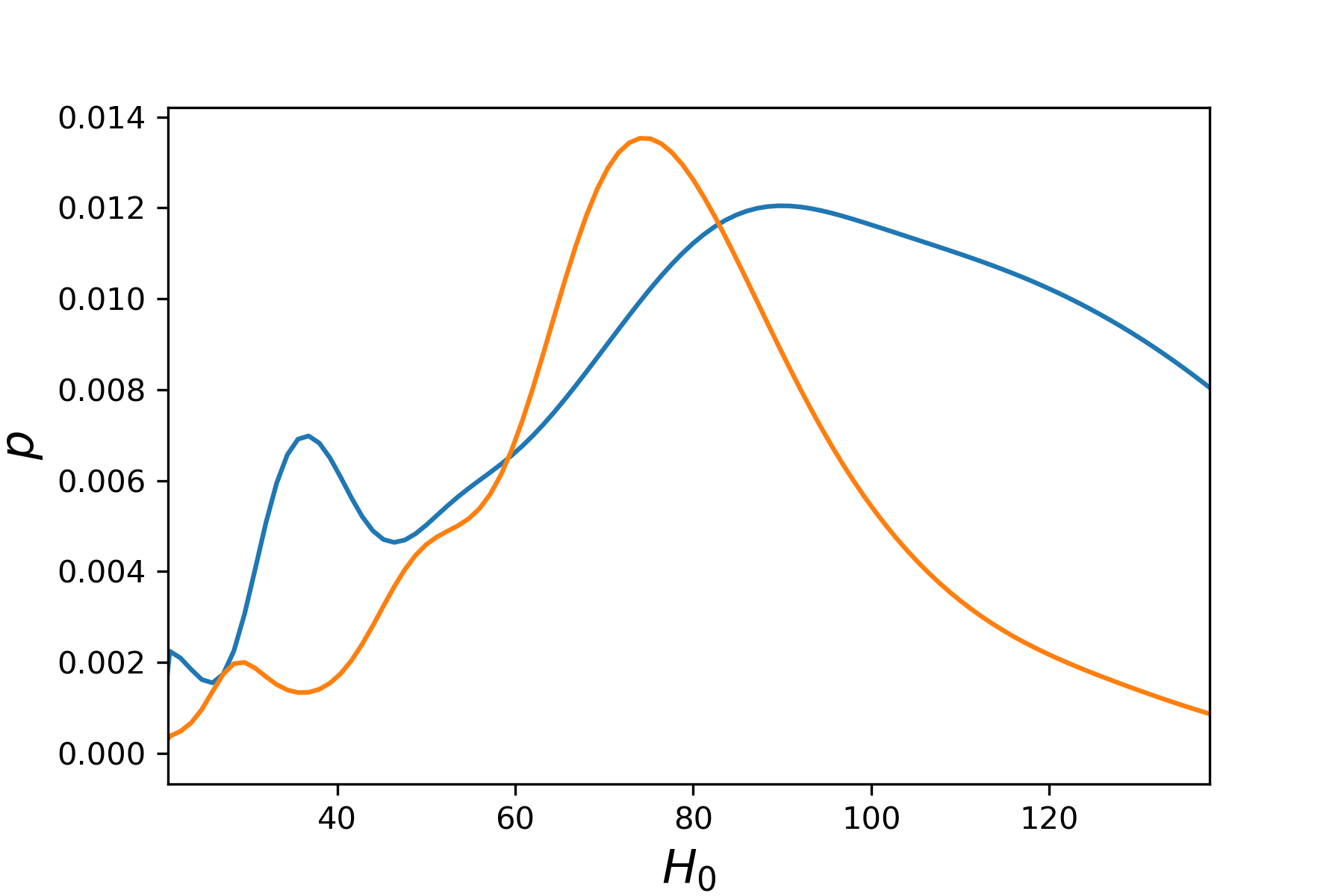} & \includegraphics[width=80mm]{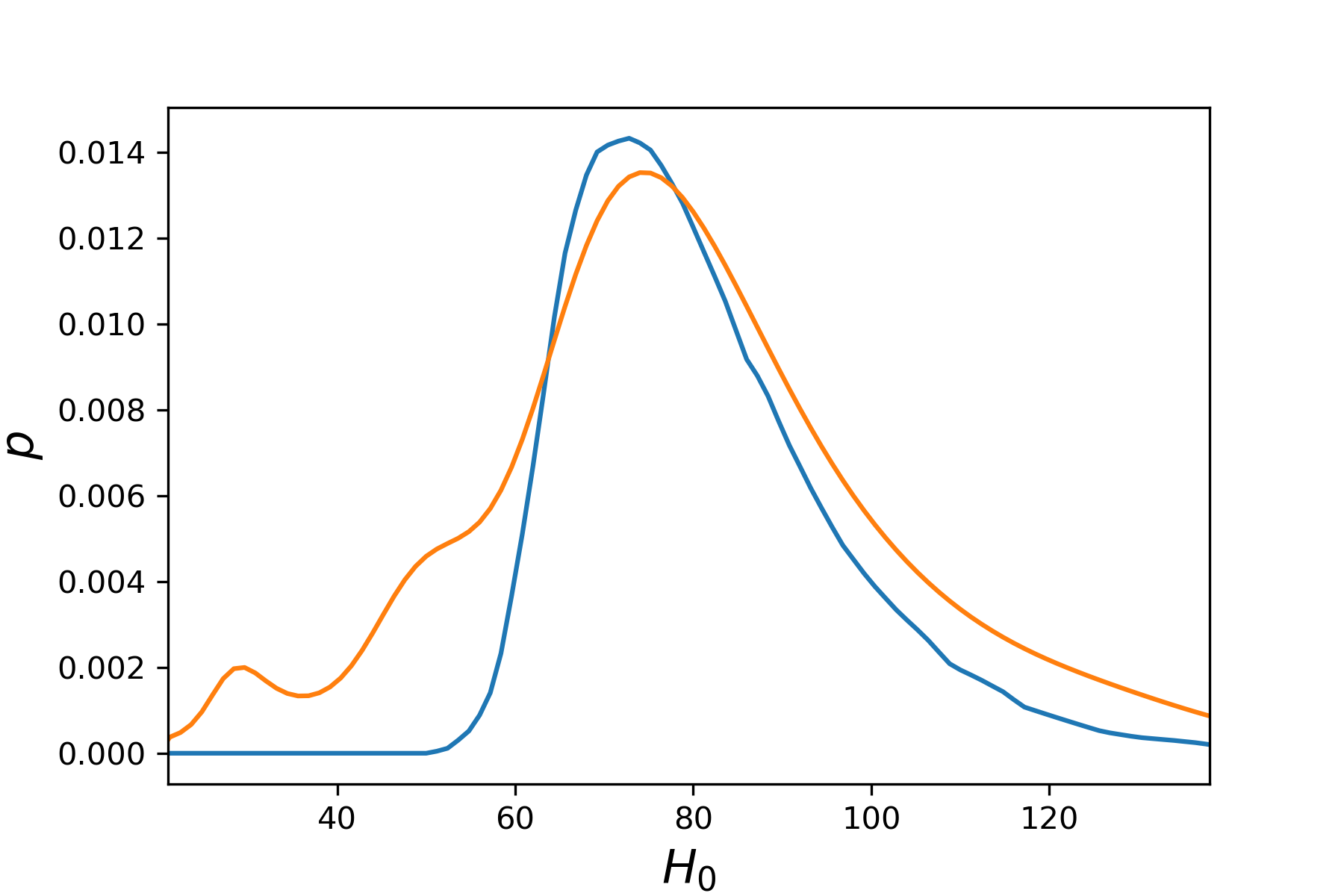}
\end{tabular}
\caption{On the left hand side we show Hubble constant posteriors for purely dark standard sirens in the CMB dipole hemisphere (blue) and the opposite hemisphere (orange). On the right hand side, we include the bright standard siren $H_0$ posterior for GW170817. The posteriors have been rescaled to aid visualisation and the Hubble constant is in the usual units, i. e. km/s/Mpc.}
\label{PDFs}
\end{figure}

This reduces us to seven events, only two of which are safely within the CMB dipole direction, namely GW190412 \cite{LIGOScientific:2020stg} in the direction of Virgo and Bo\"{o}tes constellations and the bright standard siren GW170817. Thus, our dark siren $H_0$ posterior in the CMB dipole direction, simply reduces to the GW190412 posterior from Figure 4 of Ref. \cite{Palmese:2021mjm}. Curiously, this posterior already possesses a well defined peak corresponding to a larger value, i. e. $H_0 \sim 90$ km/s/Mpc. In Figure \ref{PDFs}, on the left hand side, we reproduce the GW190412 $H_0$ posterior, which is the only event in the CMB dipole hemisphere, and combine the remaining $H_0$ posteriors from exclusively dark siren events in the opposite direction. While the errors are understandably large, it is intriguing that the maximum value of the $H_0$ posterior is larger in the CMB dipole direction. On the right hand side of the same plot, we include the bright standard siren event, which simply reverses the conclusion, but the posterior in the CMB dipole direction is dominated by GW170817. Obviously, given the poor statistics, and the fact that the dark siren events are almost exclusively oriented away from the CMB dipole, \textit{cf.} strong lensing time delay observations \cite{Krishnan:2021dyb}, it is premature to assign any statistical significance to such a small sample. That being said, since anisotropy claims in a \textit{coherent} direction have emerged \cite{Watkins:2008hf, Kashlinsky:2008ut, Lavaux:2008th, magoulas, Howlett:2022len, Migkas:2020fza, Migkas:2021zdo, Singal:2011dy, Rubart:2013tx, Tiwari:2015tba, Bengaly:2017slg, Singal:2021crs, Singal:2021kuu, Siewert:2020krp, Secrest:2020has, Krishnan:2021jmh, Luongo:2021nqh}, while GW technology is still in its infancy, it is worth highlighting the potential for GW events to weigh in on the debate surrounding the anisotropic Universe, provided suitably localised events can be found.     

\section{Conclusions}
In recent years observations have emerged hinting at an anisotropic Universe. If these observations are of physical origin, with enough precision one should not expect to find a unique value for the Hubble constant. Interestingly, existing results at low ($z \lesssim 0.05$) \cite{McClure:2007vv}, intermediate \cite{Migkas:2020fza, Migkas:2021zdo} and higher redshifts \cite{Krishnan:2021jmh, Luongo:2021nqh} now point to a potential variation of $H_0$ on the sky, at least within the flat $\Lambda$CDM framework. The goal of this note is to flag an opportunity for suitably localised future GW events to confirm this anisotropy, of course assuming it is real. Throughout, we have stressed that the local Universe out to $\sim 200$ Mpc is not an FLRW Universe, and if anything, is already anisotropic \cite{Hoffman:2017ako}. Thus, bright standard sirens \cite{LIGOScientific:2017adf} can be expected to recover this anisotropy with enough observations.

Concretely, we have decomposed the dark siren $H_0$ posteriors from Ref. \cite{Palmese:2021mjm} in hemispheres along the CMB dipole direction. In contrast to EM events, which are well localised, we rejected two events on the grounds that they potentially straddled the boundary of the hemispheres. This left six dark siren events, only one of which was in the CMB dipole direction. Curiously, we noted that the maximum of the $H_0$ posterior in the CMB dipole direction, which simply corresponded to the event G190412, was larger in line with results across strong lensing time delay \cite{Krishnan:2021dyb}, Type Ia SN \cite{Krishnan:2021jmh} and QSOs/GRBs \cite{Luongo:2021nqh}. However, once the bright standard siren event GW170817 was folded into the analysis, this reversed the conclusion. Overall, the errors are consistent with an isotropic Universe, but this exercise can be repeated as GW statistics improve. If the anisotropy is substantiated by both EM and GW events in time, it not only may explain CMB anomalies \cite{Eriksen:2003db, Hansen:2008ym, Planck:2019evm}, but raises a host of interesting theoretical questions \cite{King:1972td, Tsagas:2009nh, Tsagas:2011wq, Campanelli:2006vb, Cea:2022mtf}, which can be tested observationally \cite{Macpherson:2021gbh, Asvesta:2022fts}.

\begin{acknowledgments}
We are grateful to Antonella Palmese and for correspondence and the authors of Ref. \cite{Palmese:2021mjm} for kindly providing their Hubble constant posteriors. We thank Stephen Appleby, Asta Heinesen, Joby Kochappan for discussion and also Lu Yin for performing some preliminary work in this direction. We thank Shahin Sheikh-Jabbari and Tao Yang for comments on a final draft. We would also like to acknowledge members of the CQUEST journal club, where some of these ideas were originally teased out. E\'OC was supported by the National Research Foundation of Korea grant funded by the Korea government (MSIT) (NRF-2020R1A2C1102899). 
\end{acknowledgments}

\nocite{*}
\bibliography{sirens}

\end{document}